\documentclass[12pt]{article}
\usepackage{amsmath}
\usepackage{amssymb}
\usepackage{amsfonts}
\usepackage{amscd}
\usepackage{amsbsy}
\usepackage{amsthm}
\usepackage{latexsym}
\usepackage{graphicx,color} %,epsf}

\def\nn{\nonumber}       %%%    nonumber
\def\n{\label}                 %%%    nonumber
\def\r{\ref}                    %%%    nonumber

\def\beq{\begin{eqnarray}}
\def\eeq{\end{eqnarray}}
\def\ln{\,\mbox{ln}\,}

\def\be{\beta}
\def\ch{\chi}
\def\ga{\gamma}
\def\de{\delta}

\def\la{\lambda}
\def\na{\nabla}

\def\om{\omega}
\def\ph{\varphi}

\def\Ga{\Gamma}
\def\De{\Delta}

\def\Si{\Sigma}

\def\QG{quantum gravity}

\begin{document}

%%%%%%%%%%%%%%%%%%%%%%%%%%%%%%
\begin{center}

{\Large \bf Scalar model of effective field theory in curved space}
\vskip 8mm

{\bf Tiago G. Ribeiro}$^{a,b}$\footnote{
E-mail address: \ tgribeiro@fisica.ufjf.br},
\  \
{\bf Ilya L. Shapiro}$^{a,c,d}$\footnote{
E-mail address: \ shapiro@fisica.ufjf.br}
\vskip 3mm

$^{a}$ {\sl Departamento de Física, ICE,
Universidade Federal  de Juiz  de  Fora,
\\
Juiz de Fora, 36036-100, Minas Gerais, Brazil}
\vskip 2mm

$^{b}$ {\sl Fundação Centro de Políticas Públicas e Avaliação da Educação (Fundação CAEd),
\\
Juiz de Fora, 36038-330, Minas Gerais, Brazil}
\vskip 2mm

$^{c}$ {\sl Tomsk State Pedagogical University, Tomsk, 634041, Russia}
\vskip 2mm

$^{d}$ {\sl Tomsk State University, Tomsk, 634050, Russia}

\end{center}
\vskip 8mm

%%%%%%%%%%%%%%%%%%%%%%%%%%%
\centerline{{\bf Abstract}}

\begin{quotation}

\noindent
We consider, in more details than it was done previously, the
effective low-energy behavior in the quantum theory of a light
scalar field coupled to another scalar with much larger mass. The
main target of our work is an IR decoupling of heavy degrees of
freedom, including in the diagrams with mixed light-heavy contents
in the loops. It is shown that the one-loop diagrams with mixed
internal lines produce an IR non-local contributions which are
exactly the same as the ones in the theory of the light scalar alone,
with the effective self-interaction which can be obtained by the
functional integration of the heavy scalar, almost neglecting its
kinetic term. The same effect takes place in curved space, regardless
of a larger amount of non-localities which show up in the effective
model.
%%%%%%%%%%%%%%%%%%%%%%%%
\vskip 2mm

%% \noindent
\vskip 2mm

%% \noindent

%% \end{abstract}
\end{quotation}

%%%%%%%%%%%%%%%%%%%%%%%%%%%%%
%%%%%%%%%%%%%%%%%%%%%%%%%%%%%
\newpage

%%%%%%%%%%%%%%%%%%%%%%%%%%%%%
%%%%%%%%%%%%%%%%%%%%%%%%%%%%%
%%%%%%%%%%%%%%%%%%%%%%%%%%%%%
\section{Introduction}

The effective approach to field theory has an utmost importance at
both classical and quantum levels. The main idea behind this approach
is that low-energy physics (infrared, or IR) may be worked with
independently on the fundamental physics at high energy scale (in
the ultraviolet, or UV). For instance,
in the IR the framework of effective models assumes that we do not
need to account the degrees of freedom present in the high energy
theories \cite{Weinberg80}. The reviews of traditional realizations
of this idea in Particle Physics can be found in \cite{manohar,pich}
and the part concerning quantum gravity was extensively discussed in
\cite{Burgess:2003jk}.

The standard effective approach to quantum gravity
(see, e.g., \cite{Burgess:2003jk}) is based on treating all higher
derivative terms as small perturbations \cite{Simon-90}. This
treatment is indeed well justified if we do not care about the
underlying fundamental theory of quantum gravity that should be
valid in the UV \cite{HD-Stab}. At the same time, in \QG \ such a
fundamental theory is not known. This becomes a problem if we
recognize that all known approaches, including string theory, have
the same level of difficulties concerning higher derivatives and
ghosts \cite{ABS,Seesaw}. Thus, it looks like we need to have
to worry about ultimate quantum gravity, even if are interested
only in the IR effective approaches.

Recently, there were two new approaches to quantum gravity
which deal directly with the problem of higher derivative ghosts.
The first one is to construct the theory polynomial in metric
derivatives (with the polynomials of the even order higher than
four) and design it to be superrenormalizable \cite{highderi}.
One can provide that this theory has only complex conjugate
unphysical states. In this case one can show that the theory has
unitary $S$-matrix when quantized within the Lee-Wick approach
\cite{LQG-D4,Modesto2016}. Another possibility to avoid the
problems caused by ghosts is to introduce the nonlocal structures
into the classical action
\cite{krasnikov,kuzmin,Tomboulis-97}. The corresponding form
factors can be fine tuned to avoid the ghost states. However, it
was shown in \cite{ContGhosts} that such a fine tuning is always
destroyed by any type of \QG \ or semiclassical corrections, and
as a result at the quantum level in this theory there is an infinite
amount of ghost-like states with complex poles.

In all known approached to the fundamental theory of \QG \ the
situation is such that the higher derivative ghosts are present, being
those degrees of freedom with real poles, tachyons or some
combinations of these two types. Looking from this perspective
on the effective approach, one of the most important questions
is what remains from the higher derivative quantum gravity in the
IR, when the massive modes are supposed to decouple, including
the complex ones. This question has been posed by one of the
present authors in \cite{Gauss,Polemic}, but the answer to this
question is not known. In the present work we start to explore it
by means of the very simple toy model which admits the desired
type of decoupling.

The model which we will deal with includes  two scalar fields with
a strong hierarchy of masses, and
was in fact explored by many authors, including at the textbook
level \cite{Ilisie}. But our purposes require more detailed analysis
at the level of effective action, that will be presented below.

The decoupling theorem plays the central role in the effective
approach \cite{stephan}. This theorem \cite{AC} states that the
contribution of the loop of a field with large mass is quadratically
suppressed at low energies. The quadratic decoupling has been
explicitly checked in the framework of semiclassical gravity
\cite{apco}, but its generalization to full quantum gravity does
not look a simple task. As far as the higher derivative quantum
gravity is concerned, the high energy theory has one of many
large-mass degrees of freedom which are supposed to decouple
in the IR. But the decoupling theorem in its original formulation
does not work in this case, because some of the diagrams include
internal lines of both light and heavy fields. The questions is what
happens with the finite part of such a mixed loop in the IR, when
the energy of the fields on the external lines of the diagram is many
orders of magnitude below the largest mass in the internal
propagators? In principle, as we already mentioned above, the
corresponding calculation for the two-scalar model under
consideration is known \cite{Ilisie}, but we shall present it in
a slightly different form and also include an external gravitational
field.

The paper is organized as follows. In Sec.~\r{s2} we formulate the
model with two scalars and derive one-loop divergences and
$\be$-functions in the minimal subtraction (MS) scheme of
renormalization. As far as the model under consideration is
superrenormalizable, these $\be$-functions are indeed exact, without
further contributions at higher loop orders. Starting from the next
section, we consider the effective approach, trying to show how the
``fundamental'' two-scalar theory with cubic interactions matches
the effective one-scalar model with quartic interaction in the IR. In
Sec.~\r{s3} we discuss this matching at the tree level. The
consideration is performed in curved space and we discuss the
subtleties which show up in this case. Sec.~\r{s4} includes
derivation of one-loop diagrams with mixed (light and heavy)
internal lines. The contents of this section is almost the same
as the previous known calculations (see, e.g., \cite{Ilisie}), but
we add explanations and details which (at least in our opinion) make
the result more clear. In Sec.~\r{s5} we describe the asymptotic
behaviour of the theory in the UV and IR limits. In the former case
one meets a perfect correspondence with the MS-scheme results of
Sec.~\ref{s2}, and in the IR we can observe how the diagram with
mixed internal contents (lines of both heavy and light scalars) boils
down to the tadpole diagram of the effective model in the IR. In
Sec.~\r{s6} the previous results are generalized to the curved
space-time, by assuming weak gravitational field and using
Riemann normal coordinates, Finally, in Sec.~\r{s7} we draw our
conclusions.

%%%%%%%%%%%%%%%%%%%%%%%%%%%%%%%%%
%%%%%%%%%%%%%%%%%%%%%%%%%%%%%%%%%
%%%%%%%%%%%%%%%%%%%%%%%%%%%%%%%%%
\section{The model and its MS-scheme renormalization}
\label{s2}

In what follows in this section we formulate the classical action,
derive one-loop (which are also exact) UV divergences and the
full set of $\be$-functions.

%%%%%%%%%%%%%%%%%%%%%%%%%%%%%%%%%
\subsection{Classical action}
\label{ss21}

Consider the two-scalar model in curved space-time,  defined by
the following action:
\beq
S[\chi,\phi^{a}]
&=&
\int d^4x\sqrt{-g}
\Big\{
\frac{1}{2}\,(\na\phi^{a})^2
-  \frac{1}{2}\,m^2\phi^{a}\phi^{a}
+ \frac{1}{2}\,\xi_1R\phi^{a}\phi^{a}
\nn
\\
&+& \frac{1}{2}\,(\na\chi)^2
-\frac{1}{2}\,M^2\chi^2+\frac{1}{2}\xi_2R\chi^2
-\frac{g}{2}\,\chi\phi^{a}\phi^{a}\Big\}.
\label{action1}
\eeq
Here $\phi^a$ is the $N$-component scalar field ($\,a=1,2,\dots,N\,$)
with a mass $m$, while $\chi$ is a simple real scalar with a mass
$M$. Furthermore,
$(\na\phi^{a})^2=g^{\mu\nu}\na_\mu\phi^a \na_\nu\phi^a$,
$(\na\chi)^2=g^{\mu\nu}\na_\mu\chi \na_\nu\chi$.
Furthermore, $\,\xi_{1,2}\,$ are nonminimal parameters of
interaction between scalars fields and the curvature scalar $R$.
Later on we shall see that the quantum arguments require
supplementing the action (\ref{action1}) with the linear terms
(\ref{lin}). In this respect the situation is similar to the one for
sterile scalar field coupled to fermions \cite{AJV}, but in our
case the role of the fermions is played by the second scalar.

%%%%%%%%%%%%%%%%%%%%%%%%%%%%%%%%%
\subsection{UV divergences}
\label{ss22}

At quantum level the action (\r{action1}) leads to the theory
with the simple structure of UV divergences. Consider first the
minimal subtraction (MS) scheme. The power counting analysis
is very simple and it shows that this theory is superrenormalizable,
such that the UV divergences can be met only in the first loop and
only in the kinetic and massive terms.

We shall use short notations
$\,\int d^nx\sqrt{-g}\equiv \int_x$, with $n=4$.
Let us derive the one-loop divergences. For this end we perform
the following shift of the fields into background and quantum
counterparts:
\beq
\phi^a
\,\,\longrightarrow\,\, \phi^a+\sigma^a,
\qquad
\chi\,\,\longrightarrow\,\,\chi+\eta.
\eeq
The one-loop calculation can be done by means of the heat-kernel
method, that requires the part of the action bilinear in quantum
fields,
\beq
&&
S^{(2)}
\,=\,
\int_x
\Big\{
-\frac{1}{2}\sigma^a\square\sigma^a
-\frac{1}{2}\eta\square\eta
-\frac{1}{2}m^2\sigma^a\sigma^a
-\frac{1}{2}M^2\eta^2
\nn
\\
&+&
\frac{1}{2}\xi_1R\sigma^a\sigma^a
+\frac{1}{2}\xi_2R\eta^2
-\frac{g}{2}\chi\sigma^a\sigma^a
-g\eta\sigma^a\phi^a\Big\}
\nn
\\
&=&
-\frac{1}{2}\int_x
\begin{pmatrix}
\sigma^a &\vdots & \eta
\end{pmatrix}
\begin{pmatrix}
\delta^{ab}\big(\square+m^2-\xi_1R+g\ch\big) & \vdots &+g\phi^a\\
\hdotsfor{3} \\
+g\phi^b &\vdots& \square+M^2-\xi_2R
\end{pmatrix}
\begin{pmatrix}
\sigma^b\\
\hdots \\
\eta
\end{pmatrix}.
\mbox{\qquad}
\eeq
We have to define the matrices
\beq
\hat{\Pi} =
\begin{pmatrix}
\delta^{ab}\big(m^2-\xi_1R+g\ch\big) & \vdots &+g\phi^a\\
\hdotsfor{3} \\
+g\phi^b &\vdots& M^2-\xi_2R
\end{pmatrix}
\qquad
\mbox{and}
\qquad
\hat{P}=\hat{\Pi}+\frac{{\hat 1}}{6}R.
\eeq
One can note that the vacuum (purely metric-dependent) terms
can be easily obtained as sum of the contributions of the two free
scalar fields $\ph^a$ and $\chi$, and in general have to interest for
us. Neglecting these terms, we meet the general expression for the
one-loop divergences
\beq
\bar{\Ga}^{(1)}_{div}
\,=\,-\frac{1}{\varepsilon}\mu^{n-4}\int_x
\textrm{Tr}\Big\{\frac{1}{2}\hat{P}^2
+\frac{1}{6}\square\hat{P}\Big\}.
%% +\textrm{vacuum terms}.
\eeq
A small algebra gives us the result
\beq
\bar{\Gamma}^{(1)}_{div}
&=&
-\frac{1}{\epsilon}\mu^{n-4}
\int_x\Big\{g^2\phi^a\phi^a+\frac{Ng^2}{2}\chi^2
+Ngm^2\chi-Ng\Big(\xi_1-\frac{1}{6}\Big)\chi R
\nn
\\
&+&
\textrm{total derivative terms}\Big\},
\label{divs}
\eeq
where we introduced a compact notation $\epsilon=(4\pi)^2(n-4)$.
%%
%%   \textcolor{blue}{\bf A definição $\tilde{\xi}_i=\xi_1-\frac{1}{6}$
%%   não está sendo usada aqui.}
%%
It is clear that, in order to achieve renormalizable theory, one has to
supplement the (\ref{action1}) with two more terms,
\beq
\De S_{lin} &=& \int_x\big(\alpha\chi+\xi_3R\chi\big).
\label{lin}
\eeq
Since these terms are linear in the heavy scalar field, they do not
affect the divergences (\ref{divs}), defined by the bilinear terms.

%%%%%%%%%%%%%%%%%%%%%%%%%%%%%%%%
\subsection{$\beta$-functions}
\label{ss23}

Using the divergences, one can define the counterterms
$\De S = - \bar{\Ga}^{(1)}_{div}$, and subsequently the
renormalized action,
\beq
S_R &=& S \,+\,\De S
\nn
\\
&=& \int_x
\Big\{
   \frac{1}{2}\big(\nabla\phi^a\big)^2
+ \frac{1}{2}\big(\nabla\chi\big)^2
+\frac{1}{2}\xi_1R\phi^a\phi^a
+ \frac{1}{2}\xi_2R\chi^2
-  \frac{g}{2}\chi\phi^a\phi^a
-  \phi^a\phi^a\Big(\frac{1}{2}m^2
-  \frac{g^2}{\epsilon}\Big)
\nn
\\
&-& \chi^2\Big(
\frac{1}{2}M^2-\frac{N}{2}\frac{g^2}{2}\Big)
+\chi\Big(\alpha+\frac{Ngm^2}{\epsilon}\Big)
+ \chi R\Big[\xi_3 - \frac{Ng}{\epsilon}
\Big(\xi_1-\frac{1}{6}\Big)\Big]\Big\}
\label{Sren}
\eeq
and require it to be equal to the bare action in $n=4$ dimension.
This condition boils down to the set of renormalization relations
for the fields and one of the masses
\beq
&&
% \frac{1}{2}\big(\phi^a\big)^2\mu^{n-4}
% = \frac{1}{2}\big(\phi^a_0\big)^2
% \,\,
% \Longrightarrow
% \,\,
\phi_0^a
=\mu^{\frac{n-4}{2}}\phi^a,
\qquad
%% \nn \\ &&
% \frac{1}{2}\chi^2\mu^{n-4}=\frac{1}{2}\chi^2_0
% \,\,\Longrightarrow \,\,
\chi_0=\mu^{\frac{n-4}{2}}\chi,
\qquad
%% \nn \\ &&
m^2-\frac{2g^2}{\epsilon}=m_0^2.
\label{chibare}
\eeq

Starting from this point, we can proceed to the derivation of the
$\be$-functions. First of all, we meet the renormalization relation
\beq
0=\mu\frac{dm_0^2}{d\mu}
=\mu\frac{dm^2}{d\mu}-\frac{2}{\epsilon}\mu\frac{dg^2}{d\mu}
\,\,\,\,\Longrightarrow \,\,\,\,
\mu\frac{dm^2}{d\mu}=\frac{2}{\epsilon}\,\mu\frac{dg^2}{d\mu}.
\label{massbare}
\eeq

On the other hand,
\beq
-\frac{g}{2}\,\mu^{n-4}\chi\phi^a\phi^a
\,=\, - \frac{g_0}{2}\,\chi_0\phi^a_0\phi^a_0
\eeq
and  using (\ref{chibare}) we get
\beq
% g\mu^{\frac{4-n}{2}}\chi_0\phi^a_0\phi^a_0
% =g_0\chi_0\phi^a_0\phi^a_0\Rightarrow
g_0\,=\,g \mu^{\frac{4-n}{2}}.
\eeq

Thus,
\beq
\mu\frac{dg_0}{d\mu}
= 0 =
\mu^{\frac{4-n}{2}}\mu\frac{dg}{d\mu}+\frac{4-n}{2}g\mu^{\frac{4-n}{2}}
\,\,\Longrightarrow \,\,
\mu\frac{dg^2}{d\mu}\,=\,(n-4)g^2.
\label{gbare}
\eeq
Now, from (\ref{gbare}) and (\ref{massbare}) we get
\beq
\mu\frac{dm^2}{d\mu}
\,=\,
%% \frac{2}{(4\pi)^2(n-4)}(n-4)g^2=
\frac{2g^2}{(4\pi)^2}.
\label{betamass}
\eeq
Similarly, one can easily derive
\beq
%% M^2_0 &=& M^2-\frac{Ng^2}{\varepsilon} \nn \\ &\Rightarrow &
%% 0=\mu\frac{dM^2}{d\mu}-\frac{N}{\varepsilon}\mu\frac{dg^2}{d\mu}\Rightarrow
\mu\frac{dM^2}{d\mu}
\,=\,
%% \frac{N}{\epsilon}(n-4)g^2=
\frac{Ng^2}{(4\pi)^2}.
\eeq
Furthermore, the renormalization relation
\beq
\chi\Big(\alpha+\frac{Ngm^2}{2}\Big)\mu^{n-4}=\chi_0\alpha_0
\eeq
leads to
\beq
0\,=\,\mu^{\frac{n-4}{2}}\Big[\frac{n-4}{2}
\Big(\alpha+\frac{Ngm^2}{\epsilon}\Big)
+\mu\frac{d\alpha}{d\mu}+\frac{N}{\epsilon}
g\mu\frac{dm^2}{d\mu}+\frac{N}{\epsilon}
\mu\frac{dg}{d\mu}m^2\Big].
\label{alren}
\eeq
Since the factor $\frac{dm^2}{d\mu}$ in (\ref{betamass}) is a
function of the coupling constant, when substituted into (\ref{alren})
it leads to higher order terms of the loop expansion and hence it
can be disregarded. Thus,
\beq
\mu\frac{d\alpha}{d\mu}
&=&
- \, \frac{n-4}{2}\alpha - \frac{Ngm^2}{2(4\pi)^2}
- \frac{Nm^2}{\epsilon}\,\frac{n-4}{2}g
\,\,\,\,\Longrightarrow \,\,\,\,
\beta_\alpha=-\frac{Ngm^2}{(4\pi)^2},
\eeq
where the limit $n\rightarrow 4$ was taken.
In a similar was we obtain, in the same limit, %%  $n\rightarrow 4$,
\beq
\beta_{\xi_3}\,=\,\frac{Ng}{(4\pi)^2}\Big(\xi_1-\frac{1}{6}\Big)
\eeq
for the new nonminimal parameter introduced in (\ref{lin}). Let us
stress that all other $\be$-functions vanish and those we derived are
exact in the fundamental model (\ref{action1}).

Thinking about the correspondence between the $\be$-functions and
non-local form factors in the finite part of effective action, it is clear
that such form factors are possible only for the massive terms
$m^2$ and $M^2$, but not for the linear terms with $\alpha$ and
$\xi_3$. The corresponding $\beta$-functions are, therefore, purely
$\overline{MS}$-based, except if we consider non-local surface terms
(see \cite{Omar-surface} for an example of the corresponding
calculations).

%%%%%%%%%%%%%%%%%%%%%%%%%%%%%%%%%
%%%%%%%%%%%%%%%%%%%%%%%%%%%%%%%%%
%%%%%%%%%%%%%%%%%%%%%%%%%%%%%%%%%
\section{Matching UV and IR at the tree level}
\n{s3}

Our main interest is
% technical, namely we want
to explore in detail the decoupling in the mixed loops. However
it is worthwhile to comment on the consistency of the toy model
under consideration.

In the theory (\ref{action1}), the potential of the scalar fields in
not bounded from below. This represents a critical drawback at
the tree level, however there are two possibilities to resolve this
issue. First of all, this theory can be also a low-energy sector of
an unknown more general model, where the potential becomes
healthy and/or the two scalar fields can be composites from some
fundamental fermions, for example.   On the other hand, one can
expect that even treating the model as fundamental, the quantum
corrections change the shape of the scalar potential and the
effective potential of the theory has well defined vacuum state.
Let us start by exploring how this problem is resolved in the IR.

We assume the hierarchy $\,m\ll M$,
such that $\chi$ is a heavy field while $\phi^{a}$ is a set of light
fields with equal masses.The idea is to work out the theory at the
IR energy scale of the order $\,m\,$ and establish an effective IR
theory of this, when only the light fields $\phi^a$ are propagating.
When the energy scale is much below $M$, the oscillations of $\chi$
are suppressed, and one can expect the low-energy action of the form
\beq
\Gamma_{eff}[\phi^{a},g_{\mu\nu}]
\,=\,S[\bar{\chi},\phi^{a},g_{\mu\nu}],
\n{IRact}
\eeq
where $\bar{\chi}$ is a particular configuration of the heavy field.

One can also assume that the loops of $\chi$ are small corrections,
this point will be further discussed below. Thus, we can disregard
the term with cubic self-interaction of this field in the Lagrangian
(\ref{action1}). Then the on-shell condition can be considered at
the tree level in the form
\beq
\left.\frac{\delta S[\chi,\phi^{a},g_{\mu\nu}]}{\delta \chi}\right|_{\rm classical}
&=&
\big(\square+M^2-\xi_2R\big)\chi+\frac{g}{2}\phi^{a}\phi^{a}=0.
\n{o.sh.}
\eeq
Thus, in the classical configuration we have, as an approximation,
\beq
\chi &=&
-\,\,\frac{g/2}{\square+M^2-\xi_2R}\,\,\phi^{a}\phi^{a}.
\n{cl}
\eeq
Replacing this solution into the action (\ref{action1}) one gets the
effective low-energy action of the light field
\beq
S_{eff}[\phi^a,g_{\mu\nu}]
&=&
\int d^4x\sqrt{-g}
\Big\{
\frac{1}{2}(\na\phi^{a})^2 - \frac{1}{2}m^2\phi^{a}\phi^{a}
\nn
\\
&+& \frac{1}{2}\xi_1R\phi^a\phi^a
+\frac{g^2}{8}\,
\phi^{a}\phi^{a}\frac{1}{\square+M^2-\xi_2R}\phi^{b}\phi^{b}\Big\}.
\n{effac1}
\eeq
The action (\r{effac1}) is non-local and resembles the one we meet in
the vacuum sector in the consideration of SSB in curved space-time
\cite{sponta}. At the same time this expression becomes local if we
make further physical assumptions\footnote{Qualitatively similar
discussion of the same subject has been given recently in
\cite{Nakonechnyi}.}.

At the energy scale which is much lower than the mass $M$ we can
assume that the this mass dominates over the derivatives of the scalar,
$| \na \phi^a | \ll | M\phi^a| $ and also over the curvature
$M^2 \gg | R|$. In this case one can expand the
Green function  in a power series
\beq
\frac{1}{\square+M^2-\xi_2R}
&=&
\frac{1}{M^2}\bigg[1-\frac{\square-\xi_2R}{M^2}
+\frac{\big(\square-\xi_2R\big)^2}{M^4}
+\mathcal{O}\Big(\frac{1}{M^6}\Big)\bigg],
\label{expansion1}
\eeq
thus up to the order $1/M^6$ the action of the effective theory of
low energies is given by
\beq
S_{eff}[\phi^{a},\,g_{\mu\nu}]
&=&
\int d^4x\sqrt{-g}\,
\Big\{\frac{1}{2}(\nabla\phi^{a})^2-\frac{1}{2}m^2\phi^{a}\phi^{a}
+ \frac{1}{2}\xi_1R\phi^a\phi^a
\nn
\\
&+&
\frac{g^2}{8M^2}\phi^{a}\phi^{a}
\bigg[1-\frac{\square-\xi_2R}{M^2}
+\frac{\big(\square-\xi_2R\big)^2}{M^4} + \,\dots\bigg]
\phi^{b}\phi^{b}\Big\}.
\n{effac2}
\eeq
In the leading order this action boils down to the standard action of
the light scalar with quartic self-interaction, the effective
tree-level Lagrangian being
\beq
\mathcal{L}_{eff}[\phi^{a},g_{\mu\nu}]
&=&
\frac{1}{2}\big(\nabla\phi^{a}\big)^2
-\frac{1}{2}m^2\phi^{a}\phi^{a}
+\frac{1}{2}\xi_1R\phi^{a}\phi^{a}
-\frac{\lambda}{4!}\big(\phi^{a}\phi^{a}\big)^2,
\label{lagragianeffect}
\eeq
and the tree-level matching of the coupling is
\beq
\lambda=-\frac{3g^2}{M^2}.
\label{matching}
\eeq
In terms of the Feynman diagrams the matching condition
% (\r{matching})
means that the propagation of the heavy field is
replaced at low energies by a point interaction as shown in
Fig.~1. This should already be expected from the
expansion of the  propagator of field $\chi$ in a series of
local operators in (\ref{expansion1}).

% \noindent
%%%%%%%%%%%%%%%%%%%%%%%%%%%%%%
% Place for the 1st Figure with Feynman diagrams
%%%%%%%%%%%%%%%%%%%%%%%%%%%%%%
% \begin{figure}[!ht]
%  \centering
\beq
\includegraphics[scale=1]{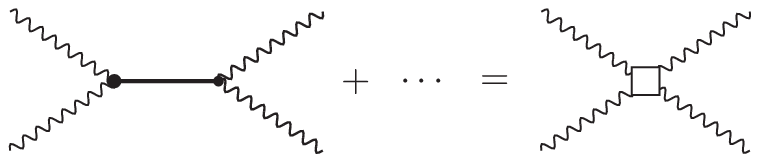}
%% \label{fig1}
\nn
\eeq
\begin{quotation}
\noindent
\textsl{Figure 1. \
Matching in terms of Feynman diagrams at the tree level.
On the left side is the diagram of the fundamental theory, dots referring
to different permutations of the momenta. At the right, the diagram
corresponds to the effective theory.}
\end{quotation}

Let us note the main difference with the IR matching in the
Standard Model of particle physics, where the diagrams with
intermediate $W$ and $Z$ bosons, in the IR, become four-fermion
interactions in the framework of the Fermi model. In the next after
the leading order - approximation in the case of (\r{effac2}) we
meet an additional curvature-dependent terms, which are not present
in the Fermi model of weak interactions. This difference shows that
for the scalars in curved space-time one has to introduce an extra
condition $M^2 \gg | R|$, in order to arrive at the effective theory
in the IR.

%%%%%%%%%%%%%%%%%%%%%%%%%%%%%%%%%
%%%%%%%%%%%%%%%%%%%%%%%%%%%%%%%%%
%%%%%%%%%%%%%%%%%%%%%%%%%%%%%%%%%
\section{One-loop calculations and effective approach}
\n{s4}

Let us explore the correspondence between the fundamental theory
(\r{action1}) and its effective IR remnant (\r{lagragianeffect}) at
the one-loop level. For the first, simpler calculation, let us start
by analysing the problem in the flat space-time, then the Lagrangian
of the complete theory is
\beq
\mathcal{L}[\chi,\phi^{a}]
=\frac{1}{2}(\partial\phi^{a})^2-\frac{1}{2}\phi^{a}\phi^{a}
+\frac{1}{2}(\partial\chi)^2
- \frac{1}{2}M^2\chi^2
- \frac{g}{2}\chi\phi^{a}\phi^{a}.
\label{fundlagrangian}
\eeq
At the tree level the matching condition (\ref{matching}).Our first
purpose is to generalize this condition to the one-loop approximation.

%%%%%%%%%%%%%%%%%%%%%%%%%%%%%
%%%%%%%%%%%%%%%%%%%%%%%%%%%%%
\subsection{One-loop corrections in full theory}
\label{ss41}

In order to analyse the IR decoupling of massive degrees of freedom
in the theory (\ref{fundlagrangian}), it is useful to consider the
diagrams that produce UV divergences and hence are responsible for
the $MS$-scheme $\be$-functions. Thus, in the fundamental theory
the corrections of interest are those for the two-point function, that
 are of the second order in the coupling constant $g$, as shown in
 Fig.~ 2.
%% \begin{figure}[!ht]
%5 \centering
\beq
\includegraphics[scale=1.1]{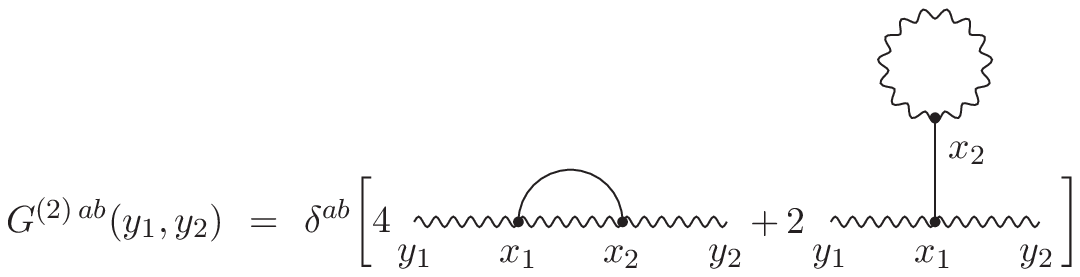}
\nn
\eeq
\begin{quotation}
\noindent
\textsl{Figure 2.
Relevant one-loop diagrams for the two-point function
of the field $\phi^{a}$ in the order $g^2$ within the fundamental
theory in the coordinate space.}
\end{quotation}
%% \label{twopointsfunction}
%% \end{figure}
These diagrams have the following analytic representation:
\beq
&&
G^{(2)\,ab}(y_1, y_2) \,=
\nn
\\
&&
=\,
\Big(\frac{-ig}{2}\Big)^2 \delta^{ab}
\int d^4x_1d^4x_2\Big[
4i\Delta_F^N(y_2-x_2)\,i\Delta_F^N(x_2-x_1)
% \nn \\ & \times &
i\Delta_F(x_2-x_1)\,i\Delta_F^N(x_1-y_1)
\nn
\\
&&
+ 2i\Delta_F^N(y_2-x_1)\,i\Delta_F^N(x_2-x_2)
\,i\Delta_F(x_2-x_1)\,i\Delta_F^N(x_1-y_1)
\Big],
\label{2pointfunc2}
\eeq
where
\beq
\Delta_F(x_1-x_2)
&=&
\int\frac{d^4p}{(2\pi)^4}\frac{e^{-ip(x_1-x_2)}}{p^2-M^2+i\epsilon}
\label{propagatorchi}
\eeq
and
\beq
\Delta^N_F(x_1-x_2)
&=&
\int\frac{d^4p}{(2\pi)^4}\frac{e^{-ip(x_1-x_2)}}{p^2-m^2+i\epsilon}.
\label{propagatorphi}
\eeq
define Feynman propagators for the $\chi$ and $\phi^{a}$
fields, respectively.

In  the  momentum  space the expression (\ref{2pointfunc2}) becomes
\beq
&&
G^{(2)\,ab}(p,-p)\,=
\nn
\\
&&
=\,
\frac{i\,\delta^{ab}}{(p^2-m^2+i\epsilon)}
\Big\{ (-ig)^2\int\frac{d^4q}{(2\pi)^4}\frac{i}{[(p-q)^2-m^2+i\epsilon]}
\frac{i}{(q^2-M^2+i\epsilon)}
\nn
\\
&&
+\,
\frac{(-ig)^2}{2}\frac{i}{(-M^2+i\epsilon)}
\int\frac{d^4q}{(2\pi)^4}\frac{i}{(q^2-m^2+i\epsilon)}\Big\}
\frac{i}{(p^2-m^2+i\epsilon)}.
\label{2pointfunc4}
\eeq

The graphical representation in the momentum space is shown in
Fig.~3.
%% \begin{figure}[!ht]
%%  \centering
\beq
\includegraphics[scale=1.1]{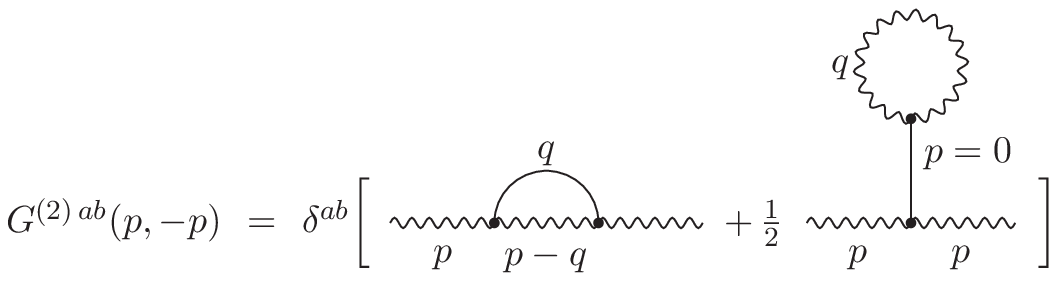}
\nn
\eeq
 \begin{quotation}
\textsl{Figure 3.\
The diagrams for the two-point function in the momentum space.}
 \end{quotation}
%% \label{twopointsfunction2}
%% \end{figure}

It proves useful to define, from (\ref{2pointfunc4}),
\beq
\Sigma^1=(-ig)^2\int\frac{d^4q}{(2\pi)^4}\frac{i}{[(p-q)^2-m^2+i\epsilon]}
\frac{i}{(q^2-M^2+i\epsilon)}
\eeq
and
\beq
\Sigma^2=\frac{(-ig)^2}{2}\frac{i}{(-M^2+i\epsilon)}
\int\frac{d^4q}{(2\pi)^4}\frac{i}{(q^2-m^2+i\epsilon)}
\eeq
so that the correction for the 2-point function is given by the expression
\beq
&&
G^{(2)\,ab}(p,-p)\,=\,
\frac{i\,\delta^{ab}}{(p^2-m^2+i\epsilon)}
(\Sigma^1+\Sigma^2)
\frac{i}{(p^2-m^2+i\epsilon)}.
\eeq
It is easy to see that the first of these quantities has logarithmic
divergence and depends on the external momentum in an essential
way. Our purpose is to verify how this expression interpolates
between UV and IR regions and what remains from its finite part
in the limit $M \to \infty$.
On the other hand, the second (tadpole) diagram is quadratically
divergent, but the dependence on the external momentum is trivial.

%%%%%%%%%%%%%%%%%%%%%%%%%%%%
%%%%%%%%%%%%%%%%%%%%%%%%%%%%
\subsection{Dimensional regularization}
\label{ss31}

Using dimensional regularization (see \cite{Leibrandt} for the
introduction), we generalize the  divergent expressions in
Eq.~(\ref{2pointfunc4}) to the integrals in $2\omega$-dimensional
Euclidean space,
\beq
\Sigma^1_{2\omega}
&=&
ig^2(\mu^2)^{2-\omega}
\int\frac{d^{2\omega}q}{(2\pi)^{2\omega}}
\frac{1}{[(q-p)^2+m^2](q^2+M^2)}.
\label{i2omega}
\eeq
where $\mu$ is a dimensional renormalization parameter.

In order to solve $\Sigma^1_{2\omega}$, we use the presentation
\beq
\frac{1}{ab}=\int_0^1\frac{dz}{[az+b(1-z)]^2}
\nn
\eeq
and perform a change of variablel $q^\prime=q-p(1-z)$, with
$dq^\prime=dq$. Thus, $\Sigma^1_{2\omega}$ becomes
\beq
\Sigma^1_{2\omega}
&=&
ig^2(\mu^2)^{2-\omega}\int\frac{d^{2\omega}q}{(2\pi)^{2\om}}
\int_0^1\frac{dz}{\big[q^2+p^2z(1-z)+(M^2-m^2)z+m^2\big]^2},
\label{i2omega1}
\eeq
where $q^\prime$ was replaced by $q$. %  for the sake of simplicity.
The integral over $q$ can be easily taken in a standard way using
spherical coordinates in the $2\omega$-dimensional  momentum
space. The  result is
\beq
\Sigma^1_{2\omega}
&=&
\frac{ig^2}{(4\pi)^2}\,\Ga(2-\omega)
\int_0^1dz\bigg[
\frac{4\pi\mu^2}{\bigl(p^2z(1-z)+(M^2-m^2)z+m^2\bigr)}
\bigg]^{2-\omega},
\label{i2omega2}
\eeq
where $\Ga(t)$ is Euler's Gamma function.

Taking the limit $\,\omega \to2$ one can use the expansions
\beq
\Ga(t)\,=\,\frac{1}{t}-\ga + \mathcal{O}(t)
\hspace{0.5cm}
\textrm{and}
\hspace{0.5cm}
x^t=e^{t\ln x}\simeq 1+t\ln x,
\label{aproximation}
\eeq
where $\gamma \approx 0,577$ is the Euler-Mascheroni constant.
Thus, we arrive at the following result
\beq
\Sigma^1
&=&
\frac{ig^2}{(4\pi)^2}\bigg\{
\frac{1}{(2-\om)} - \ga
- \int_0^1dz \ln\Big[\frac{p^2z(1-z)+(M^2-m^2)z+m^2}{4\pi\mu^2}\Big]
\bigg\}.
\label{i2omega3}
\eeq
The integral in  the last expression can be easily solved in the form
\beq
\int_0^1dx\,\ln\Big[1+\frac{4x(1-x)}{a}+4bx\Big]
&=&-\,2 +\frac{(1-ab)}{2}\log(1+4b)
\nn
\\
&+&
\frac{A}{2}\,\log \left[ \frac{(A+1)^2 -a^2b^2}{(A-1)^2-a^2b^2}\right],
\label{int}
\eeq
with the restrictions and notation
\beq
a>0, \quad  b>- 1/4
\quad
\mbox{and}
\quad
A=\sqrt{\big(1+ab\big)^2+a}.
\nn
\eeq

In order to use the result (\r{int}), we set
\beq
a=\frac{4m^2}{p^2}
\qquad
\mbox{and}
\qquad
b=\frac{M^2-m^2}{4m^2}.
\n{ab}
\eeq
Finally, the first part of the one-loop contribution to the $2$-point
function is
\beq
\Sigma^{1}
&=&
\frac{ig^2}{(4\pi)^2}\Big\{\frac{1}{\varepsilon}
+ 2+ \log\Big(\frac{\mu^2}{m^2}\Big)
-\frac{1-ab}{2}\log\bigg(\frac{M^2}{m^2}\bigg)
\nn \\
&-&
\frac{A}{2}\,\log \left[ \frac{(A+1)^2 -a^2b^2}{(A-1)^2-a^2b^2}\right],
%% \log \Big[\frac{(A+1)^2-(ab)^2} {(A-1)^2-(ab)^2}\Big]\Big\},
\label{correction}
\\
\mbox{where}
&&
\frac{1}{\varepsilon}=\frac{1}{2-\om}-\ga +\log4\pi.
\label{div}
\eeq
In the limit $\,\omega \to2$ the result for $\Sigma^{1}$ has
divergent and finite parts. It is easy to check that the divergent
part corresponds to the result (\ref{divs}) in the $\phi^a\phi^a$
sector. At the same time, the dependence on the external momenta
in the finite part is rather complicated and indicates a non-locality
of the effective action.

Working out the second  integral in (\ref{2pointfunc4}) gives,
in the same limit,
\beq
\Si^{2}
&=&
\frac{ig^2}{2}\frac{(\mu^2)^{2-\omega}}{M^2}
\int\frac{d^{2\omega}q}{(2\pi)^{2\omega}}\,\frac{1}{q^2+m^2}
%% \nn \\ &=&
\,=\,
\frac{ig^2\,m^2}{2(4\pi)^2\,M^2}
\Big(\frac{4\pi\mu^2}{m^2}\Big)^{2-\om}\Ga(1-\om).
\label{sigmatwo}
\nn
\\
&=&
-\,\frac{ig^2}{2(4\pi)^2}\frac{m^2}{M^2}\,\Big[\frac{1}{\varepsilon}+ 1 + \log\Big(\frac{\mu^2}{m^2}\Big)\Big],
\label{tadpole}
\eeq
disregarding ${\cal O}(2-\om)$ terms. There is  only a local contribution,
as it has to be for the tadpole.

%%%%%%%%%%%%%%%%%%%%%%%%%%%%%%%%
\section{Asymptotic behavior}
\label{s5}

Now we are in a position to explore both high-energy and
low-energy regimes in the two-point function. Let's take the UV
limit ($p^2\rightarrow\infty$) in the expression  (\ref{correction}).
The relations $\,p^2 \gg m^2\,$ and $\,p^2 \gg M^2\,$ result in
\beq
&&
\Sigma^1_{\rm {UV}}(p^2\rightarrow\infty)
\nn
\\
&&
=\,\frac{ig^2}{(4\pi)^2}\Big\{\frac{1}{\varepsilon}
+ 2 - \log \Big(\frac{p^2}{\mu^2}\Big)
- \frac{m^2}{p^2}\Big[1+\log\Big(\frac{p^2}{m^2}\Big)\Big]
-\frac{M^2}{p^2}\Big[1+\log\Big(\frac{p^2}{M^2}\Big)\Big]\Big\},
\mbox{\qquad}
\label{Sigma}
\eeq
where all lower order terms are omitted. The logarithmic terms in the
form factor are proportional to the divergence, as it should be in
the UV. Furthermore, it is easy to check that the divergent term
exactly correspond the result in Eq.~(\ref{divs}). This
correspondence has a relevant consequence. Let us remember that
the theory is superrenormalizable and that Eq.~(\ref{divs}) give all
UV divergences which we can meet in all loop orders. This means
that higher loop corrections to  (\ref{Sigma}) are also finite and,
moreover, they do not have higher order logarithmic corrections.
Thus, Eq.~(\ref{Sigma}) is the leading contribution not only at the
one-loop level, but also non-perturbatively.

In the IR we assume $p^2 \ll M^2$ in the expression (\ref{correction}).
In the leading order in $p^2$ this yields
\beq
\Sigma^1_{\rm{IR}}(M^2\gg p^2)
\,=\,
\frac{ig^2}{(4\pi)^2}\Big\{\frac{1}{\varepsilon}+ 1
+ \log\Big(\frac{\mu^2}{M^2}\Big)
+ \frac{m^2}{M^2}\log\Big(\frac{m^2}{M^2}\Big)
- \frac{1}{2}\frac{p^2}{M^2}\Big\}.
\label{ircorrection}
\eeq
It is easy to see that there is no nonlocal part with a logarithmic form
factor. Thus, the diagram with mixed (light and heavy fields) internal
lines boils down to the tadpole-type contribution in the IR.

%%%%%%%%%%%%%%%%%%%%%%%%%%%%%%%%%%%
%%%%%%%%%%%%%%%%%%%%%%%%%%%%%%%%%%%
\subsection{Matching with IR at the one-loop level}
\label{ss51}

Once we know the one-loop behavior of the fundamental theory in the
IR, it is possible to establish the correspondence between this result
and that obtained from an effective theory, taking into account only
the quartic interaction of the field $\phi^{a}$ at the one-loop level.
At the first stage, we shall disregard the correction (\ref{tadpole}) for
the fundamental theory, since this contribution is independent of the
momentum. The graphical representation of the one-loop matching
is shown in Fig.~4.
\beq
\includegraphics[scale=1.1]{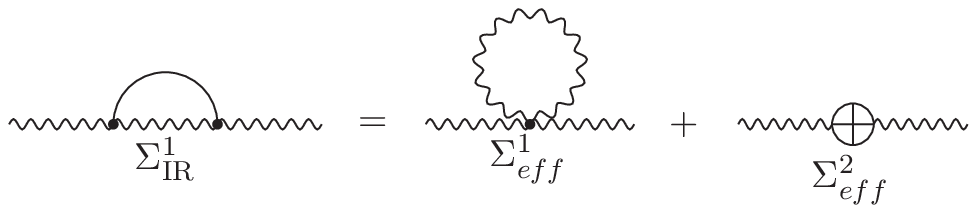}
\nn
\eeq
\begin{quotation}
 \noindent
 \textsl{Figure 4. \ Illustration of the one-loop match in the IR
 between  fundamental theory (left side of equality) and the
 effective theory with the four-scalar interaction (right side of equality).}
 \end{quotation}
%% \end{figure}

%%%%%%%%%%%%%%%%%%%%%%%%%%%%%

On the {\it l.h.s.}, $\Sigma^1_{IR}$ is the correction to the
propagator of the $ \phi^{a}$ field in the fundamental theory in
the IR limit (\ref{ircorrection}). On the {\it r.h.s.},
$\,\Sigma_{eff}^1\,$ is the one-loop correction for the propagator
in the effective theory, whereas $\,\Sigma_{eff}^2\,$ is an
additional term (established below) representing the difference
between the one-loop correction of the fundamental theory in the
low energies and the correction to one-loop of the effective theory
of low energies.
%% red
%% \textcolor{red}{\bf Tiago, como definir este termo?}
%% \textcolor{blue}{\bf Ele é definido pelos novos termos em (50) o que significa
%% $\Sigma^2_{eff}=\Sigma^1_{\rm{IR}}-\Sigma^1_{eff}$.}

The effective theory in the IR, at the tree level corresponds to
a quartic interaction of $N$ scalars fields (\ref{lagragianeffect})
in the flat space,
\beq
\mathcal{L}_{eff}[\phi^{a}]
\,=\,
\frac{1}{2}(\partial\phi^{a})^2
- \frac{1}{2}m^2\phi^{a}\phi^{a}
- \frac{\lambda}{4!}(\phi^{a}\phi^{a})^2,
\label{lagraneff}
\eeq
where $\lambda=-\frac{3g^2}{M^2}$. The one-loop contribution
in this effective theory can be easily calculated to give
\beq
\Sigma_{eff}^1
\,=\,-\,\frac{3ig^2 m^2}{2(4\pi)^2M^2}\Big[\frac{1}{\varepsilon}
+ 1 + \log\Big(\frac{\mu^2}{m^2}\Big)\Big],
\label{oneloopeff}
\eeq
with $\frac{1}{\varepsilon}$ being defined in (\ref{div}).

The additional term $\Sigma^2_{eff}$ is obtained by inserting new
coefficients in the effective Lagrangian
\beq
\mathcal{L}_{eff}[\phi^{a}]
\,=\,\frac{1}{2}(1+C_{\phi})(\partial\phi^{a})^2
- \frac{1}{2}(m^2+C_{m^2})\phi^{a}\phi^{a}
- \frac{\lambda}{4!}(\phi^{a}\phi^{a})^2,
\label{lageff}
\eeq
providing the vertices
\beq
\Sigma^2_{eff}\,=\,-iC_{m^2}+ip^2C_{\phi}.
\eeq

These coefficients should not be confused with the counterterms. Their
role is not to remove divergences from theory, but to ensure that at low
energies both effective and fundamental theory lead to identical results,
as shown below.

The divergences in (\ref{ircorrection}) and (\ref{oneloopeff}) can be
removed by a suitable renormalization. Since the renormalization scale
is arbitrary, we are interested only in the finite part of these corrections.
The one-loop matching is achieved by
\beq
\Sigma^1_{\rm{IR}}=\Sigma^1_{eff}+\Sigma^2_{eff},
\label{mathcing1loop}
\eeq
remembering that we are interested in the low energy regime of
the fundamental theory.

The equality (\ref{mathcing1loop}) leads to the following values,
\beq
C_{m^2}
\,=\,
-\frac{g^2}{(4\pi)^2}\Big[\Big(1+\frac{m^2}{M^2}\Big)
\Big(1 + \log\frac{\mu^2}{M^2}\Big) +\frac{m^2}{2M^2}
\Big(1+\log\frac{\mu^2}{m^2}\Big)\Big]
\label{cm1}
\eeq
and
\beq
C_{\phi}\,=\,- \, \frac{g^2}{2(4\pi)^2M^2}.
\eeq
These results for $C_{m^2}$ and $C_{\phi}$ show how the effective
theory differs from the fundamental theory of low energies. It is important
to note that these two terms are momentum-independent, and therefore
can be compensated by the change in renormalization condition.

It is easy to see from the expression of the tadpole (\ref{tadpole}),
that if $\Sigma^2$  is added on the left side of (\ref{mathcing1loop})
the last term of $C_{m^2}$ cancels and we arrive at
\beq
C_{m^2}
\,=\,-\, \frac{g^2}{(4\pi)^2}\Big(1+\frac{m^2}{M^2}\Big)
\Big(1 + \log\frac{\mu^2}{M^2}\Big).
\label{resutoneloop}
\eeq

When $M^2\rightarrow\infty$  the results for $C_{m^2}$ and $C_{\phi}$
confirm the decoupling theorem. In the IR, the difference between
fundamental and effective theories is reduced by renormalization of
UV divergences by irrelevant local counterterms, and in the terms
proportional to the inverse of the square of the mass of the heavy field.

From (\ref{lageff}) and the results obtained by the IR matching is
possible to note that the effective theory can be written as
\beq
\mathcal{L}_{eff}\,=\,\frac{1}{2}C_1(\partial\phi^a)^2
- \frac{1}{2}C_2\phi^a\phi^a-\frac{1}{4!}C_3(\phi^a\phi^a)^2,
\eeq
where the coefficients $C_1$, $C_2$ and $C_3$ depend on the
parameters of the fundamental theory valid in any energy scale.
In general, these coefficients can be constructed order by order in
the loop expansion, ensuring that the results of the two theories are
equivalent at low energies.

%%%%%%%%%%%%%%%%%%%%%%%%%%%%%%%%%%%%
%%%%%%%%%%%%%%%%%%%%%%%%%%%%%%%%%%%%
\section{Decoupling in a weak gravitational field}
\label{s6}

In this section we will generalize the previous considerations to the
same theory in a curved space. As usual, the derivation of non-local
form factors requires that the metric corresponds to the almost flat
geometry. Then, the external metric can be treated as a small
perturbation. Alternatively, one can make an expansion of the
$2$-point function in normal coordinates \cite{bunch} and directly
arrive at the formally covariant result for the form factor. In this
section we shall follow this approach and use the expansion in the
linear order in curvature tensor components.

Starting from the first term of (\ref{2pointfunc2}), we write the
2-point function in the form
\beq
G^{(2)ab}(y_1,y_2)
&=&
\delta^{ab}\int d^4x_2 \sqrt{- g(x_2)}\int d^4x_1 \sqrt{- g(x_1)}
\nn
\\
&\times &
G^N(y_2,x_2) \Big[(-ig)^2 G^N(x_2,x_1)G(x_2,x_1)\Big] G^N(x_1,y_1)
\label{ER}
\eeq
where $G^N(x,y)=i\Delta_F^N(x-y)$ and $G(x,y)=i\Delta_F(x-y)$ are the
flat-space Green functions for the light and heavy fields, respectively.

After Wick rotation to Euclidean space, this 2-point function can be
written as
\beq
G^{(2)ab}(y_1,y_2)
\,=\,
- i\de^{ab}\int_{x_2}\int_{x_1} %%  d^4x_2\sqrt{g(x_2)}
%% \int d^4x_1 \sqrt{g(x_1)}
\,G^N(y_2,x_2)\,\Sigma^{1,R}\,G^N(x_1,y_1),
\label{curved-0}
\eeq
where
\beq
\Sigma^{1,R}=ig^2G^N(x_2,x_1)G(x_2,x_1).
\eeq
Let us note that $\Sigma^{1,R}$ includes the product of the two
propagators that appear in the loop. These propagators can be expanded
in terms of normal coordinates in a curved space according to the
well-known result
%% (4.20)
of \cite{bunch}, which in $2\omega$-dimensions provides
\beq
\Sigma^{1,R}
&=&
ig^2
\int\frac{d^{2\omega}q}{(2\pi)^{2\omega}}\,e^{iq(x_2-x_1)}
G(q,m)
\int\frac{d^{2\omega}p}{(2\pi)^{2\omega}}\,e^{ip(x_2-x_1)}
G(p,M),
\label{ER1}
\eeq
where $\Sigma^{1,R}=\Sigma^{1,R}(k)$, \ $q + p = k$,
\beq
G(q)
\,=\,
\frac{1}{q^2+m^2}+\frac{\tilde{\xi_1} R}{(q^2+m^2)^2}
-\frac{2}{3}\frac{R_{\alpha\beta}q^\alpha q^\beta}{(q^2+m^2)^3}
+ \mathcal{O}(q^{-5})
\label{G}
\eeq
is the standard expansion of the propagator \cite{bunch}, and we are
using notations
\beq
\tilde{\xi_1}=\frac{1}{3}-\xi_1
\hspace{1cm} \rm{and} \hspace{1cm}
\tilde{\xi_2}=\frac{1}{3}-\xi_2.
\label{eq1}
\eeq
Let us note that the expansion of the vertices in normal coordinates
in Eq.~(\ref{ER1}) is not necessary, as explained below.
%% red
%% \textcolor{red}{\bf Falta colocar expansao de vertices, eu acho.}
%% \textcolor{blue}{\bf Aqui não precisa colocar expansão do vértice.
%% As correções são proporcionais a uma função delta dirac que leva
%%  as expressões da teoria fundamental para um único vértice no limite IR.}

Consider the dimensional regularization of the integrals in
Eq.~(\ref{ER1}), using the transformations for the integrals with
vector symmetry described in the books \cite{Ilisie, schwartz}. Our
final purpose is to evaluate these integrals in the IR limit with
$\,m^2\ll M^2$ and $p^2\ll M^2$. Setting $x_2-x_1=y$, after
some algebra we obtain
\beq
I_1
&=&
\int\frac{d^{2\omega}k}{(2\pi)^{2\omega}}e^{iky}
\int\frac{d^{2\omega}p}{(2\pi)^{2\omega}}
\,\,\frac{1}{[(p-k)^2+m^2](p^2+M^2)}
\nn
\\
&=&
\frac{1}{(4\pi)^2}\int\frac{d^4k}{(2\pi)^4}e^{iky}
\Big[\frac{1}{\varepsilon}+ 1 + \log\Big(\frac{\mu^2}{M^2}\Big)
+ \frac{m^2}{M^2}\log\Big(\frac{m^2}{M^2}\Big)
-\frac{1}{2}\frac{k^2}{M^2}\Big].
\label{eq10}
\eeq
This is the same expression as (\ref{ircorrection}), that means for
$R = 0$ we recover the flat space result. Other integrals correspond
to the dependence on the curvature,
\beq
I_2
&=&
\tilde{\xi_1}R\int\frac{d^{2\om}k}{(2\pi)^{2\omega}}
e^{iky}\int\frac{d^{2\omega}q}{(2\pi)^{2\omega}}
\frac{1}{[(q-k)^2+M^2](q^2+m^2)^2}
\nn
\\
&=&
\frac{\tilde{\xi_1}R}{(4\pi)^2}\int\frac{d^4k}{(2\pi)^4}
\frac{e^{iky}}{M^2}\Big[-1-\log\Big(\frac{m^2}{M^2}\Big)\Big]
\nn
\\
I_3
&=&
-\frac{2}{3}R^{\alpha\beta}\int\frac{d^{2\omega}k}{(2\pi)^{2\omega}}
e^{iky}\int\frac{d^{2\omega}q}{(2\pi)^{2\omega}}
\frac{q_\alpha q\beta}{(q^2+m^2)^3[(q-k)^2+M^2]}
\nn
\\
&=&
-\frac{1}{6}\frac{R}{(4\pi)^2}
\int \frac{d^4k}{(2\pi)^4}\frac{e^{iky}}{M^2}
\Big[-\frac{3}{2}-\log\Big(\frac{m^2}{M^2}\Big)\Big]
\nn
\\
I_4
&=&
\tilde{\xi_2}R\int\frac{d^{2\omega}k}{(2\pi)^{2\omega}}e^{iky}
\int\frac{d^{2\omega}q}{(2\pi)^{2\omega}}
\frac{1}{(q^2+m^2)[(q-k)^2+M^2]^2}
\nn
\\
&=&
\frac{\tilde{\xi_2}R}{(4\pi)^2}\int\frac{d^4k}{(2\pi)^4}e^{iky}\frac{1}{M^2}
\nn
\\
I_5
&=&
-\frac{2}{3}R^{\alpha\beta}\int\frac{d^{2\omega}k}{(2\pi)^{2\omega}}
e^{iky}\int\frac{d^{2\omega}q}{(2\pi)^{2\omega}}
\frac{(k_\alpha -q_\alpha)(k_\beta -q_\beta)}{(q^2+m^2)[(q-k)^2+M^2]^3}
\nn
\\
&=&
-\frac{1}{6}\frac{R}{(4\pi)^2}\int \frac{d^4k}{(2\pi)^4}e^{iky}
\frac{1}{2M^2},
\eeq
where we have set $\omega=2$, since there are no divergences.

Disregarding the $k^2$-term in Eq.~(\ref{eq10}),
%% red
%% \textcolor{red}{\bf O que significa isso?}
%% \textcolor{blue}{\bf Se eu não fizer essa consideração (62) entra
%% com um termo proporcional a $k^2$ e a integral não cai direto em uma delta.}
the combination of these integrals gives
\beq
\Sigma^{1,R}_{IR}
&=&
\delta(y)\frac{ig^2}{(4\pi)^2}\Big\{
\frac{1}{\varepsilon}+\log\Big(\frac{1}{M^2}\Big)
+\frac{m^2}{M^2}\log\Big(\frac{m^2}{M^2}\Big)
\nn
\\
&+&
\frac{R}{M^2}\Big[\Big(\xi_1-\frac{1}{6}\Big)
\log\Big(\frac{m^2}{M^2}\Big) + \xi_1 - \xi_2 + \frac{1}{6}
\Big]\Big\}.
\label{curved}
\eeq

An important observation is in order. The expression (\ref{ER1}) does
not include the expansion of the vertex in normal coordinates. The
reason is that the expansion of the vertex comes from the factors of
$\sqrt{g}$ in the interaction terms. In one of the points (e.g. $x_1$)
the metric is flat, such that  $\sqrt{g(x_1)}=1$, and in another
point the expansion boils down to the factor of $\sqrt{g}$ in the
final expression for $\Sigma^{1,R}_{IR}$ in Eq.~(\ref{curved}),
since this is a local expression that has an extra factor of
$\,\de_c(x_2-x_1)$. This one should be a covariant delta function in
normal coordinates, absorbing the whole factor of $\sqrt{g}$, that
comes from the second vertex. The delta function deletes one of the
integrals in Eq.~(\ref{curved-0}), such that the results becomes a
local expression.

In the IR limit, the 2-point function in (\ref{curved-0}) can be
written in the form
\beq
G^{(2)ab}(y_1,y_2)
\,=\,-\,i\delta^{ab}\int d^4x \sqrt{g(x)}
\; G^N(y_2,x)\;\Sigma^{1,R}_{IR} \;G^N(x,y_1).
\label{curvedonepoint}
\eeq
%% where $\Sigma^{1,R}_{IR}$ is now the same as (\ref{curved})
%% but without the delta function.
where  $\Sigma^{1,R}_{IR}$ no longer proportional to the delta
function. It is easy to see that the expression (\ref{curvedonepoint})
has UV divergences only in the flat-space sector, while the terms with
scalar curvature are finite. This output is in a perfect correspondence
with the covariant calculation in Sec.~\ref{s2}.

In order to compare the two approaches to the description of the
IR, let us now consider an effective theory with quartic interaction
in curved space. Such a the theory of the field $\phi^{a}$ leads to
the one-loop contribution
\beq
G^{(2)ab}_{eff}(y_1,y_2)
\,=\,-\,
i\de^{ab}\int d^4x \sqrt{g(x)} \; G^N(y_2,x) \Big[-\frac{i\la}{2}
\,G^N(x,x)\Big]G^N(x,y_1).
\label{effcorrection}
\eeq
This expression can be directly compared to (\ref{curvedonepoint}),
to show that both have the same structure in terms of propagators.
This comparison is possible by the fact that in low energies the
fundamental theory has only one vertex.

The matching at the one-loop level in the curved space can be
done in pretty much the same way as in flat space,
\beq
\Sigma^{1,R}_{IR} = \Sigma_{eff}-iC_{m^2}+iC_R
\label{curvedmatching}
\eeq
Here $\Sigma_{eff}$  is defined from (\ref{effcorrection}) as
\beq
\Sigma_{eff}\,=\,-\frac{i\la}{2}\,G^N(x^\prime,x^\prime).
\eeq
Using the expansion in normal coordinates, we get
\beq
\Sigma_{eff}
\,=\,-\,\frac{i\lambda}{2}\int \frac{d^{2\om}p}{(2\pi)^{2\om}}
\,\bigg\{
\frac{1}{p^2+m^2}
+\frac{\tilde{\xi_1}R}{(p^2+m^2)^2}
-\frac{2}{3}\frac{R_{\alpha\beta}p^\alpha p^\beta}{(p^2+m^2)^3}
+\mathcal{O}(p^{-5})\bigg\},
\eeq
that leads to the result
\beq
\Sigma_{eff}
&=&
\frac{i\lambda}{2(4\pi)^2}\bigg\{
m^2\Big[\frac{1}{\varepsilon}
+ 1 + \log\Big(\frac{\mu^2}{m^2}\Big)\Big]
+ \Big(\xi_1-\frac{1}{6}\Big) R\Big[\frac{1}{\varepsilon}
+ \log\Big(\frac{\mu^2}{m^2}\Big)\Big]\bigg\}.
\eeq
Note that, according to Eq.~(\ref{matching}), even in curved space
we have $\,\la=-\frac{3g^2}{M^2}$. Thus, the flat part of the IR
matching condition in  (\ref{curvedmatching}) is satisfied with the
$C_{m^2}$ from (\ref{cm1}). On the other hand, there is an
additional  matching condition in the first order in curvature,
%\beq
%C_R\,=\,\frac{g^2}{(4\pi)^2}\frac{R}{M^2}
%\Big[\tilde{\xi}_1 \log\Big(\frac{\mu^2}{M^2}\Big)
%+ \frac{1}{2}\tilde{\xi}_2
%\log\Big(
%\frac{\mu^2}{m^2}\Big)-\Big(\tilde{\xi}_2-\tilde{\xi}_1 -\frac16
%\Big)\Big].
%\eeq
\beq
C_R
\,=\,
\frac{g^2}{(4\pi)^2}\frac{R}{M^2}
\Big[\Big(\xi_1-\frac{1}{6}\Big)\log\Big(\frac{\mu^2}{M^2}\Big)
+ \frac{1}{2}\Big(\xi_1-\frac{1}{6}\Big)
\log\Big(\frac{\mu^2}{m^2}\Big)
+ \big(\xi_1 - \xi_2\big) + \frac{1}{6} \Big].
\eeq

As in the flat space, we can consider the correction of tadpole
to the fundamental theory, given by the second term of
(\ref{2pointfunc2}) in curved space. In Euclidean space it is
\beq
G^{(2)ab}_{tadpole}(y_1,y_2)
\,=\,-\,i\de^{ab}\int_{x_2} \int_{x_1}
\; G^N(y_2,x_1)\Big[\Sigma^{2,R}\Big]G^N(x_1,y_1).
\label{tadpolecurved}
\eeq
with
\beq
\Sigma^{2,R}
&=&
\frac{ig^2}{2}G(x_2,x_1)G^N(x_2,x_2)
\nn
\\
&=&
\frac{ig^2}{2}\int\frac{d^4p}{(2\pi)^2}
\frac{e^{ip\cdot(x_2-x_1)}}{M^2}
\Big[1+\mathcal{O}\Big(\frac{p^2}{M^2}\Big)\Big]G^N(x_2,x_2)
\mbox{\qquad}
\nn
\\
&=&
\frac{ig^2}{2M^2}\,\delta_c(y)
\int\frac{d^{2\om}q}{(2\pi)^{\om}}
\Big[\frac{1}{q^2+m^2}+\frac{\tilde{\xi_1}R}{(q^2+m^2)^2}
-\frac{2}{3}\frac{R_{\alpha\beta}q^\alpha q^\beta}{(q^2+m^2)^3}
+\mathcal{O}(q^{-5})\Big].
\mbox{\qquad}
\eeq
In this expression, the covariant delta function emerges in the low
energy limit, when expanding the propagator of the heavy field in
powers of $\frac{p^2}{M^2}$ and disregard higher orders in the
expansion.

The tadpole part gives the contribution
\beq
\Sigma^{2,R}\,=\,-\,
\frac{g^2}{2(4\pi)^2}\Big\{\frac{m^2}{M^2}
\Big[\frac{1}{\varepsilon}+1+\log\Big(\frac{\mu^2}{m^2}\Big)\Big]
+ \Big(\xi_1-\frac{1}{6}\Big)\frac{R}{M^2}\,
\Big[\frac{1}{\varepsilon}+\log\Big(\frac{\mu^2}{m^2}\Big)
\Big]\Big\}.
\eeq
Adding this result to the {\it l.h.s.} of (\ref{curvedmatching}), the flat
part of the matching at one-loop level leads to the expression
(\ref{resutoneloop}), while the curvature-dependent part gives
\beq
C_R=\frac{g^2}{(4\pi)^2}\frac{R}{M^2}
\Big[\Big(\xi_1-\frac{1}{6}\Big)\log\Big(\frac{\mu^2}{M^2}\Big)
+ \big(\xi_1 - \xi_2\big) + \frac{1}{6}\Big].
\eeq

%%%%%%%%%%%%%%%%%%%%%%%%%%%%%%%%%%
%%%%%%%%%%%%%%%%%%%%%%%%%%%%%%%%%%
%%%%%%%%%%%%%%%%%%%%%%%%%%%%%%%%%%
\section{Conclusions}
\label{s7}

Using a very simple model with two scalar fields, we explored the
behavior of the diagrams with mixed internal lines. To some extent
the results are not new (see, e.g., \cite{Ilisie}), but we made the
calculations keeping in focus the effective action approach,  the
relevant problem of decoupling of higher derivatives in quantum
gravity \cite{Polemic} and considered in full details the matching
between UV and IR, including in the weak external gravitational
field.

The main output of our investigation is that the contribution of the
self-energy type, one-loop diagram with one internal line of the light
field and another one of the field with the large mass, in the far IR
boils down to the tadpole contribution, that does not produce a
non-local form factor. In the toy model under consideration this
means that the self-energy diagram in the ``fundamental'' model
with two types of scalars produce a standard non-local form factor
with the logarithmic asymptotic behavior in the UV, but in the IR
there is no relevant form factor and the results is essentially the
same tadpole-type contribution that one can get in the effective
low-energy model with a single-type light scalar field. The same
qualitative situation holds in a weak gravitational field. Indeed,
due to the superrenormalizable nature of the fundamental model,
the logarithmic asymptotic behavior in the UV is not possible for
the curvature-dependent terms. However, in the IR we observe a
perfect matching between effective and fundamental models, that
confirms the main results of our work.

From the gravitational perspective, the massive fields are ghosts
and tachyons that are present in the higher derivative versions of
quantum gravity. In this respect the important question is whether
the IR effective theory is always the quantum general relativity,
or it can be some other, e.g., non-local model, as it was discussed
in \cite{Polemic}. Making a ``continuation'' of our present result
implies that one can expect that the mixed-content diagrams become
irrelevant in the IR. Then the transferred momentum is the unique
IR regulator and this means that the quantum general relativity is
expected to be a universal model of IR quantum gravity, as it was
expected in the paper by Donoghue \cite{don94} and many
consequent works (see, e.g., the reviews \cite{don2}). Indeed, this
kind of conclusion should be seen as a well-motivated conjecture,
and its verification would be an interesting work to be done.

%%%%%%%%%%%%%%%%%%%%%%%%%%%%%%%%%%
%%%%%%%%%%%%%%%%%%%%%%%%%%%%%%%%%%
%%%%%%%%%%%%%%%%%%%%%%%%%%%%%%%%%%
\section{Note Added}
\label{s8}

Regardless we mainly use the theory (\ref{action1}) as a toy
model for quantum gravity, it is worthy to discuss whether this model
can be independently applied to the description of some physical
phenomena. The idea looks attractive, because \ \textit{i)} it is
superrenormalizable and therefore does not need complicated
nonperturbative treatment. On the other hand, the standard four-scalar
model emerge naturally in the IR, as we have seen in Sec.~\ref{s2}.
This situation should enable one to avoid the well-known difficulties
with the stability of Higgs potential in the UV (see e.g.
\cite{Riotto-2009} and \cite{Strumia-2012}).

Unfortunately, this apparently nice plan meets two serious obstacles.
First of all, the classical potential of the scalar fields in the theory
(\ref{action1}) is not bounded from below and, therefore, its quantum
mechanical formulation meets a fundamental difficulty. Moreover,
since the theory is superrenormalizable, the loop contributions
to the quantum effective potential
% \footnote{We will not bother the
% reader with the corresponding calculations at one loop, but they
% confirm this assumption.}
only enhance logarithmically the massive
terms. Thus, even in the non-perturbative effective potential one
should expect the directions that make the potential unbounded, that
confirm the tree-level verdict. Second, if we assume that the scalar
fields $\phi^a$ and $\chi$ couple to fermions, the nice features
of the model immediately break down, as the renormalizability
should require $\,(\phi^a\phi^a)^2$, $\,\phi^a\phi^a\chi^2$ and
$\,\chi^4\,$ terms to be introduced, and then the theory is not
superrenormalizable anymore.

Thus, there is no much chances to transform our model into the
base of the fundamental theory behind the Standard Model of Particle
Physics (SM). At the same time, the simplified versions of the model
(\ref{action1}) are known as useful tools in cosmology. E.g., in
the well-known papers \cite{Boyanovsky-2012,Lankinen-2018}
(see further references therein, also the recent work \cite{Erdem})
the scalar field  $\chi$ describes the Bose-Einstein condensate
of some more fundamental fields. The considerations in these
works involves not only tree-level, but also the loop effects. Thus,
it is possible that the consistent formulation of the model in curved
space which we presented in sections~\ref{s2} and \ref{s3},
may be useful for
further developments of this approach, the same concerns its IR
quantum behavior, that we discussed in the subsequent sections.

Finally, let us mention the possibility that the large-mass field 
$\chi$ in our physically motivated toy model may turn out to be 
a natural concept in the models of composite Higgs.
Up to some extent, this can be a particle physics version of the
cosmological applications considered in \cite{Boyanovsky-2012}.
Once again, in this case it may be useful to have a consistent
formulation of the model in curved space, e.g. because it may put
an additional restrictions of the heavy scalar field as a composite
object coming from some fundamental fermions beyond the SM,
for example.

%%%%%%%%%%%%%%%%%%%%%%%%%%%%%%%%%%
%%%%%%%%%%%%%%%%%%%%%%%%%%%%%%%%%%
%%%%%%%%%%%%%%%%%%%%%%%%%%%%%%%%%%
\section*{Acknowledgements}
\noindent
Authors are grateful to Bla\v{z}enka Meli\'{c} and Oleg Antipin from
RBI/Zagreb for useful discussions and clarifications concerning the
effective approach in particle physics. The work of I.Sh. was
partially supported by Conselho Nacional de Desenvolvimento
Cient\'{i}fico e Tecnol\'{o}gico (CNPq) under the grant 303635/2018-5
and by Funda\c{c}\~{a}o de Amparo \`a Pesquisa de Minas Gerais
(FAPEMIG) under the project APQ-01205-16.

%\section*{References}
%%%%%%%%%%%%%%%%%%%%%%%%%%%%%%%%%%%%
%%%%%%%%%%%%%%%%%%%%%%%%%%%%%%%%%%%%
%%%%%%%%%%%%%%%%%%%%%%%%%%%%%%%%%%%%

%\end{multicols}

\end{document}